# Bi-static Radar Cross Section Test Method by Using Historic Marconi Set-up and Time Gating

Yousef Azizi, Mohammad Soleimani, Seyed Hasan Sedighi, and Ladislau Matekovits, *Senior Member, IEEE*

*Abstract*— In this paper, a low-cost, simple and reliable bi-static Radar Cross Section (RCS) measurement method making use a historic Marconi set-up is presented. It uses a transmitting (Tx) antenna (located at a constant position, at a reference angle of θ = 0º) and a receiver (Rx) antenna (mounted on a moveable arm calibrated in the azimuthal direction with an accuracy of 0.1º). A time gating method is used to extract the information from the reflection in the time domain; applying time filter allows removing the antenna side lobe effects and other ambient noise. In this method, the Rx antenna (on the movable arm) is used to measure the reflected field in the angular range from 1º to 90º of reflection from the structure (printed PCB) and from the reference configuration represented by a ground (GND) plane of the same dimension. The time gating method is then applied to each pair of PCB / GND measurements to extract the bi-static RCS pattern of the structure at a given frequency. Here comparison of measurement results carried out at 18 GHz and 32 GHz with simulation indicates the successful performance of the proposed method. It can be used as a low-cost, reliable and available option in future measurement and scientific research.

*Index Terms*—Radar Cross Section, Marconi Set-up, Time Gating.

## I. INTRODUCTION

RCS measurement is one of the essential requirements in the field of telecommunication engineering, of which evaluation is always controversial due to the existence of measurement errors, environmental noise, post processing requirement to extract results, and expensive automation measuring equipment [1]. What is known as the RCS was first discussed in the field of military issues [2]-[4]. However, this concept was introduced in non-military applications at 1965 [5]. The RCS measurement techniques have been well presented in [6], also. Performing RCS measurement on a laboratory scale requires hardware such as shielded chamber, absorbers, Tx/Rx antennas, RF signal sources and receivers [7]. Also, extracting and processing measurement information to eliminate background noise, antenna side lobe effects and other environmental reflection factors require using of numerical ways and accessories such as time gating which is well discussed for example in [8]. The RCS measurement methods for static and dynamic facilities were presented in [9] and [10], respectively. During the last decades, many methods such as method of moment (MOM), geometrical theory of diffraction (GTD) and physical optic (PO) have been proposed in [11]. Moreover, many numerical tools such as finite element method (FEM), boundary element method and other software packages were developed to easily calculate the RCS of structures [12]. Bi-static RCS measurement by using some special equipment (synchronized Tx/Rx, source and receiver equipment, moving rails, etc.) is expensive, and therefore few laboratories have the ability to measure it [13, 14]. In order to solve the problem of expensive automation set-ups, using Marconi training set-up can be effective as a simple and reliable solution in the static RCS measurement. The Marconi set-up, which is used as one of the teaching aids in laboratories, has a movable arm and an angled calibrator (protractor) whose rotation rate can be adjusted and controlled [15, 16]. However, the automatic or manual solutions for RCS measurement are possible, also.

In this paper, a simple, cost-effective, and reliable numerical bi-static RCS measurement method is presented to extract the RCS results by mounting Marconi set-up inside of the anechoic chamber, and using the time gating method. In this measurement, the Tx antenna is placed in the front of the PCB/GND of the sample structure, and the Rx antenna is moved manually at the same time on the rotating-angular calibrated arm to extract the reflection information at all angles from 1° to 90° for different frequencies. Then, the PCB/GND measurement information at each angle is post-processed, by converting them from frequency to time domain by using Inverse Fast Fourier Transform (IFFT). The data are then time filtered and re-converted to the frequency domain by using FFT. The procedure is applied for both the PCB and GND reflection results at each angle and the bi-static pattern is extracted at each frequency, consequently. Since the measurement methods and numerical tools of CST Studio software are in good agreement with the IEEE 1502 recommended [12] (known as the main reference for RCS measurement), the measurement results of proposed method

Manuscript received **** accepted *****.
Y. Azizi and M. Soleimani are with the Department of Electrical Engineering, Iran University of Science and Technology (IUST), Tehran, Iran (e-mail: yousefazizi@elec.iust.ac.ir; soleimani@iust.ac.ir).
S. H. Sedighy is with the School of New Technologies, Iran University of Science and Technology, Tehran, Iran (e-mail: sedighy@iust.ac.ir)
L. Matekovits is with the Department of Electronics and Telecommunications, Politecnico di Torino, Corso Duca degli Abruzzi 24, I-10129 Turin, Italy (e-mail: ladislau.matekovits@polito.it). He is also with Istituto di Elettronica e di Ingegneria dell'Informazione e delle Telecomunicazioni, National Research Council of Italy, 10129 Turin, Italy and with Department of Measurements and Optical Electronics, University Politehnica Timisoara, 300006 Timisoara, Romania.

are compared and validated with the simulation ones. Comparison of simulation and measurement results of a sample structure indicates that this simple, reliable and cost-effective test method could be used as a measurement method.

## II. BI-STATIC TEST METHOD

The Marconi test set-up as one of the old equipment in communication laboratory is shown in Fig. 1a.

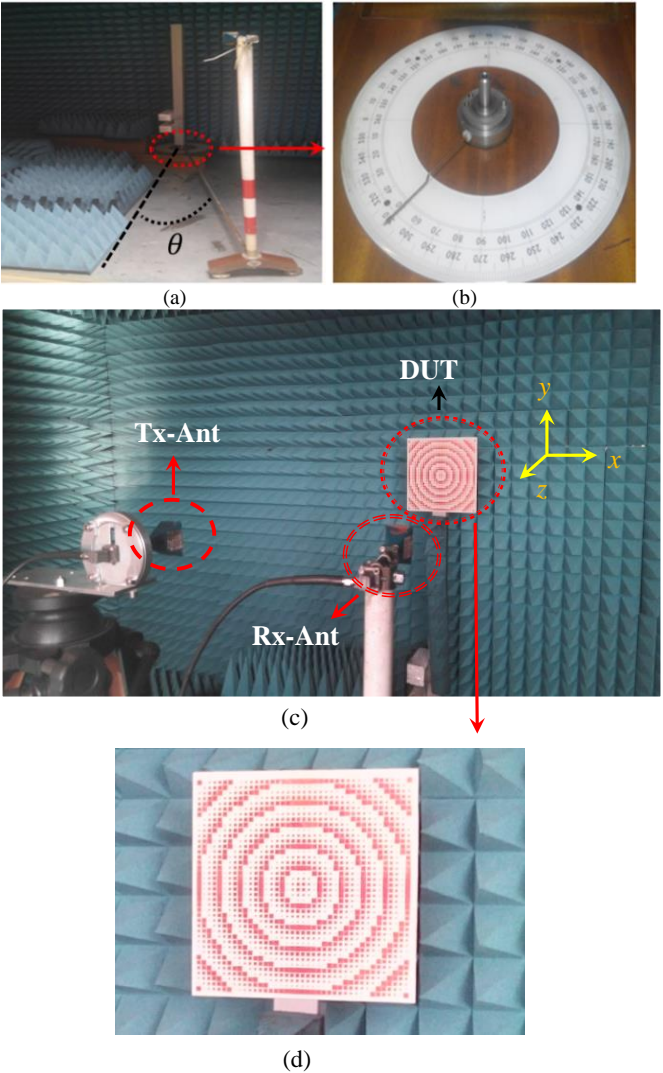

Fig. 1. Photograph of the measurement set-up and DUT.
a) Schematic of Marconi set-up in anechoic chamber;
b) Angular calibrated surface of the Marconi set-up;
c) Bi-static set-up using Marconi set-up, Tx/Rx antenna and DUT;
d) Sample metasurface to be tested

This set-up has a central stand for placing the device under test (DUT) on it, a rotating and calibrated arm where the amount of rotation can be seen and controlled (Fig. 1b, c). In order to measure bi-static RCS, the Tx antenna is placed in front of the target at the same height of the DUT, while Rx antenna is placed on the rotating arm still at the same height to cover any angle for which one to measure, store and process the reflection signal from the target. The prototype metasurface structure in Fig. 1 (d) is bi-statically RCS tested with this method to evaluate the performance of the proposed method. The metasurface under test has of $250 \times 250$ mm$^2$ dimension that is printed on a single layer grounded RO4003 with 1.6 mm thickness (Fig.1 d).

In order to measure the bi-static pattern of the sample structure by using Marconi set-up, it is necessary to fix the Tx antenna in the front of target, and then measure the reflection from PCB / GND of DUT by the Rx antenna mounted on a movable arm at any angle. The Rx antenna is rotated manually by using the calibrated screen of the set-up angle meter (Fig. 1 b) to minimize the angular error. Since the structure of the sample metasurface has symmetry with respect to x / y axes (horizontal and vertical in the figure), it is appropriate to measure the reflection from the PCB / GND of the metasurface in the angle range of 1° - 90°.

## III. POST PROCESSING PROCEDURE

After measuring the reflection from PCB / GND at all desired angles (θ = 1° ~ - 90°) by using time gating method, the PCB/ GND reflection signals (both real and imaginary parts) at any specific angle θ are converted from the frequency domain to the time domain. The discreet IFFT equation is used as [17]

$$X_k = \sum_{n=0}^{N-1} x_n e^{-i2\pi kn/N}, k = 0,...,N-1 \quad (1)$$

where $x_n$ is the $n^{th}$ value of the reflection signal (real and imaginary) in the frequency domain, $N$ is the number of frequency samples and $X_k$ is the $k$th value of the signal in the time domain. Then, by applying a suitable time filter (Kaiser) in the time domain, noise and other existing environmental reflections are removed. At this stage, only the PCB / GND reflection signals remain, and by FFT, the reflection rate from the PCB / GND is calculated in terms of frequency. The FFT equation that used is as [17]

$$x_n = \frac{1}{N}\sum_{n=0}^{N-1} X_k e^{i2\pi kn/N}, k = 0,...,N-1 \quad (2)$$

where $X_k$ is the $k$th value of the signal in time domain, $N$ is the number of time steps and $x_n$ is the $n$th value of the signal at frequency domain (real and imaginary). This coordinate loop shown in Fig. 2 flowchart is used to extract the PCB / GND reflection at all angles (θ = 1° - 90°). In order to apply the measurement flowchart presented in Fig. 2 and measure the bi-static RCS of the metasurface at 18 GHz and 32 GHz, N5227A PNA Network Analyzer is used. The test is performed in one of the anechoic chamber at Polytechnic of Turin, from 10 to 40 GHz (30 GHz bandwidth) covered by 3 sets of TX/Rx antennas, at 1601 frequency points at each frequency band, and 100 Hz PNA resolution bandwidth. The frequency bandwidth, steps and resolution bandwidth are chosen so that the reflected signal from PCB / GND in the time domain may be detected [1]. It should also be noted that the radial distance between Tx/Rx antenna to PCB / GND is equal to 2.5 meters that gives rise to a high spatial attenuation, especially at high frequency.

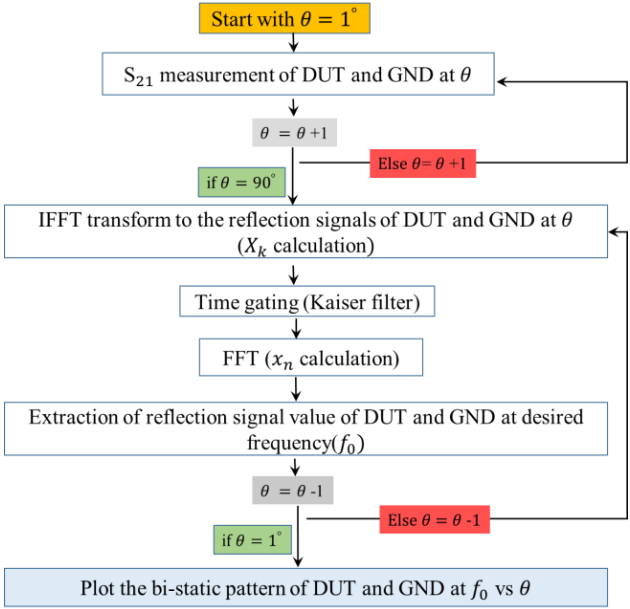

Fig. 2. Flowchart of RCSR measurement using Marconi set-up and time gating method

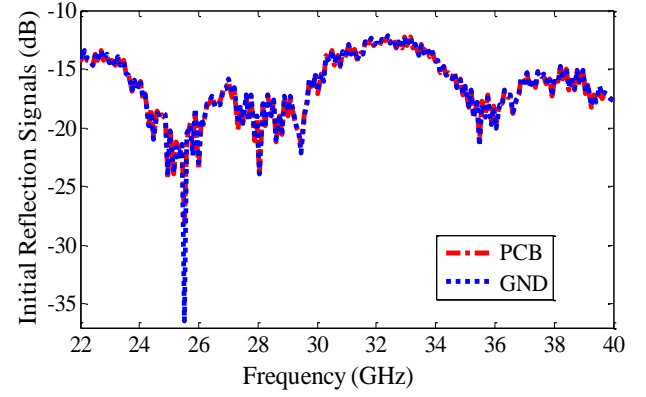

(a)

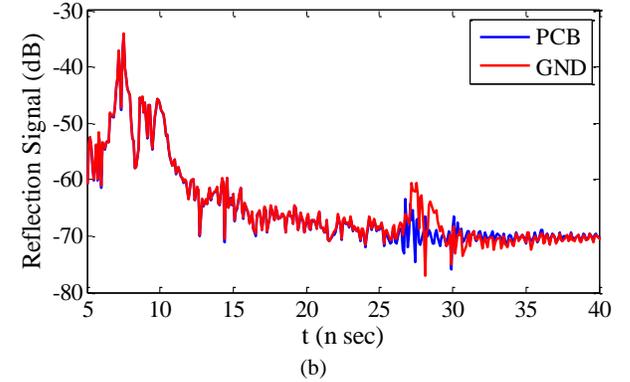

(b)

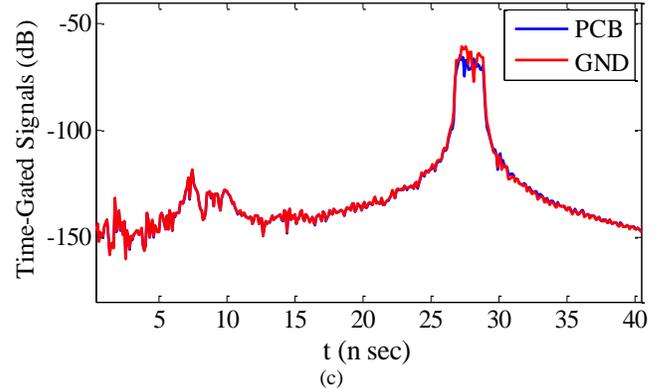

(c)

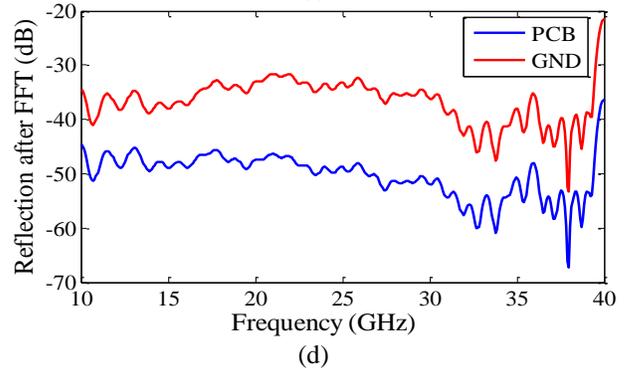

(d)

By starting from θ = 1°, the reflection from PCB / GND is measured at all 1° -90° angles, respectively. It should be noted that the reflection measurement from DUT is done by the network analyzer at the frequency domain, and in this case the reflection of both PCB and GND in the frequency domain has the same power range as depicted in Fig. 3 (a). In this case, it is impossible to distinguish between the reflection signal of PCB and GND due to the similar amplitude of their reflection power. At this stage, the distinction between the reflection signals and consequently the time gate required to apply the time filter is determined by applying IFFT and transmitting the PCB and GND reflection signals from the frequency domain to the time domain. Figure 3 (b) shows the PCB / GND reflection at θ = 1° in the time domain (which correspond to the 22-40 GHz TX/Rx antenna). It can be seen that there is a difference in the reflection amplitude of PCB and GND in the time interval of 27-29 nsec. In fact, applying a proper time filter to the reflection signals in this time interval reduces the background noise, side lobe signals (especially at 6-10 nsec range that is related to the side lobes) and reflection from other environmental factors with more than 80 dB. Figure 3 (c) shows the PCB / GND reflection after applying the high order (400) Kaiser time filter. The difference between the reflection signal of PCB and GND in the period of 27-29 nanoseconds can be seen in Fig. 3 (c). By applying the appropriate time filter (Kaiser) as well as FFT to the signals of this time period, the reflection of PCB and GND in the frequency domain and subsequently RCSR can be calculated.

All post processing steps, i.e. IFFT, FFT and Kaiser filter calculation are implemented in Matlab software.

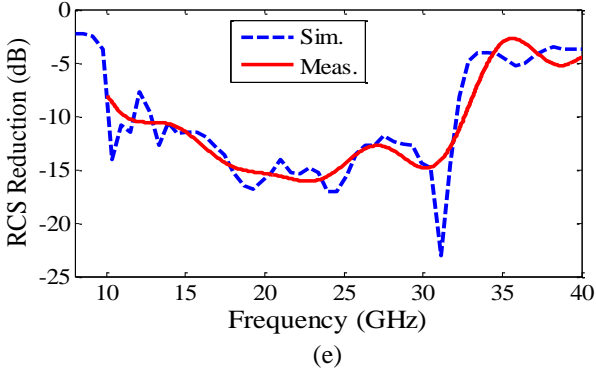

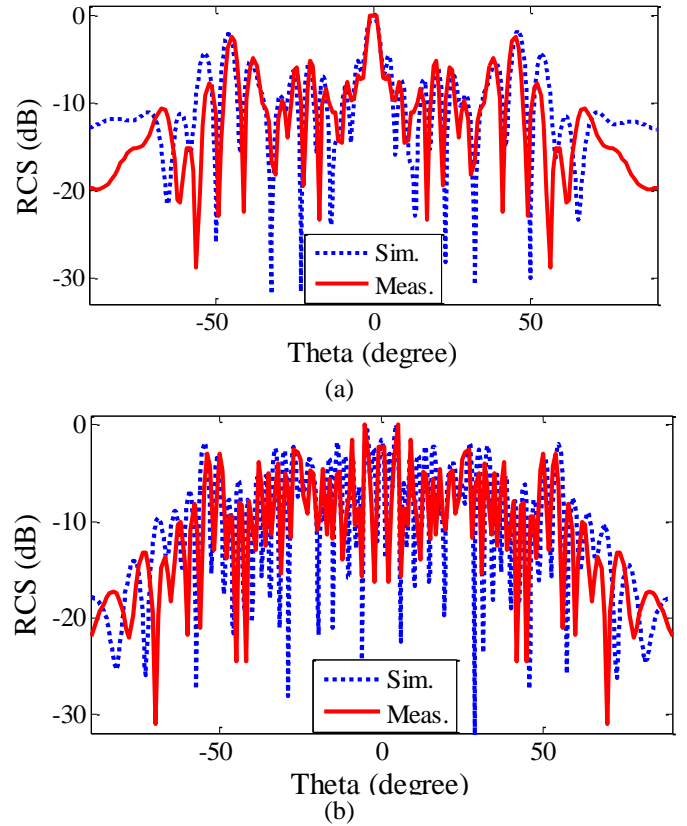

Fig. 3. Time domain reflection results of the PCB/GND and RCS results
a) Initial reflection signal of PCB/GND in frequency domain
b) Time domain measurement results of PCB/GND
c) PCB/GND reflection results after applying Kaiser time gating filter
d) Reflection signal of PCB/GND after FFT
e) RCSR measurement and simulation of the sample metasurface at θ = 1°

In the next step, the PCB / GND reflection for the considered frequency band (10-40 GHz) at θ = 1° is obtained by applying the IFFT which is plotted in Fig. 3 (d). As shown in this figure, the reflection from the PCB surface is less than the GND reflection in the entire frequency range, 10 GHz to 40 GHz. Simply put, the metasurface under test has broadband RCSR performance. At this stage, the signal difference between PCB and GND is extracted as a measured RCSR, which is plotted in Fig. 3 (e).

The simulation and measurement results of Fig. 3 (e) are in good agreement with each other and there are small differences (lower than 1.5-dB) between simulated and measured results at 18 GHz and 32 GHz, which indicates the efficiency of the proposed test method. Comparison of the simulation results provided by the full-wave CST software (which has analytical tools and numerical methods in accordance with the IEEE 1502 standard) with the measured ones proves the accuracy and reliability of the proposed method.

## IV. BI-STATIC RCS PATTERN EXTRACTION

In the following, the reflection values are extracted from other angles as shown in Fig. 2. By performing the above procedure for all of the other angles and extracting the reflection values from PCB/GND at 18 GHz and 32 GHz (at any angle), the bi-static RCS pattern of sample metasurface was plotted in Fig.4 (a) and (b), respectively.

For better present and validate the test method, the simulated and measured normalized PCB/GND RCS patterns are able to compare with each other due to the symmetry of the structure, the measurement results from θ = 1° to 90° are mirrored to the θ = -1° to -90°. Note that the measurement results for θ = 0° is not applicable (because of the Tx and Rx antenna should share the same location); when using two Tx/Rx antenna the minimum measureable angle is 1°.

Fig. 4. Measurement and simulation bi-static results of the proposed metasurface at a) 18 GHz and b) 32GHz

Comparison of measurement and simulation results at 18 GHz shows that there is a good agreement between them in the angular range of 1° - 60°. According to Fig. 4 (a), there is a difference up to several dB between the simulation results and the measurement at angles greater than 60°. The reason for this error can be due to the errors in the environmental factors such as set-up vibration, openness in the chamber gate, and errors related to the construction of metasurfaces and calibration of test equipment. In general, there is a good similarity and agreement between the measurement results with the proposed method and the simulation results at 18 GHz, which is due to the following reasons: (i) accurate angular calibration that is achieved using Marconi set-up and allows accurate measurement of reflection at any angle. (ii) reflection noise effect elimination from the environment by the time filter, which reduces a significant part of the environmental measurement error (advantage of time gating method). In a similar way, a comparison between the measurement and simulation RCS pattern at 32 GHz shows that the proposed method has good accuracy. In this case, the number of resonances in the RCS pattern of the structure has increased, which can be seen in the measurement results. In a similar way, a comparison between the RCS pattern measurement and simulation results at 32 GHz proves that the proposed method has good accuracy. In this case, the number of resonances in the RCS pattern of the structure has increased, which can be seen in the measurement results, also.

Unlike the simulation results at 18 GHz frequency and 60-90° incident waves, there are good agreement between the measurement and simulation results.

## V. Conclusion

A low-cost, simple, and reliable bi-static test method was proposed using a calibrated Marconi scientific-educational test set-up to perform the measurements and post-processed by time-gating method ( using FFT, time filtering and IFFT) used to extract scattering pattern of objects and structures. This method can be used as a suitable alternative to expensive bi-static measurement methods that require expensive hardware and software equipment. Using time gate and applying appropriate time filtering (in this case Kaiser filter was used) leads to eliminate background and environment reflection, also. The performance of the proposed method is confirmed by testing a sample metasurface and comparing it with the simulation results. The sample metasurface has dimensions of $250 \times 250$ mm$^2$, which is printed on the Rogers 4003 substrate with a thickness of 1.6 mm. Comparison of simulation and measurement results shows that the proposed method has a potential to be widely used in laboratory and scientific applications by students and researchers. Also, the comparison between bi-static measurement results and simulation results (which has analytical tools and numerical methods in accordance with the IEEE 1502 standard) confirms the validation of the proposed method.

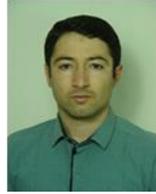

Yousef Azizi was born in Kermanshah, Iran, in 1989. He received his B.Ss and M.Sc. degree all in Electrical Engineering from Urmia University (UU) and Iran University of Science and Technology (IUST), Iran, in 2013 and 2016, respectively. He is currently working toward the Ph.D. degree in Communication Engineering in ISUT, Tehran, Iran. His major research interests are the design of Metasurface, Superstrate Antenna and Modulated surface for Radar Cross Section Reduction (RCSR).

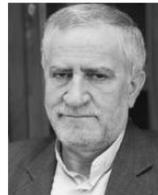

Mohammad Soleimani received the B.Sc. degree in electrical engineering from the University of Shiraz, Shiraz, Iran, in 1978, and the M.Sc. and Ph.D. degrees from Pierre and Marie Curie University, Paris, France, in 1981 and 1983, respectively. He is currently a Professor with the School of Electrical Engineering, Iran University of Sciences and Technology, Tehran, and serves as the Director with the Antenna and Microwave Research Laboratory. He has also served in many executive and research positions. He has authored and coauthored 19 books (in Persian) and more than 200 journal and conference papers. His research interests include electromagnetics and high-frequency electronics and antennas.

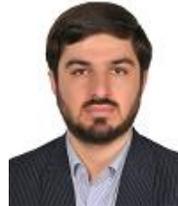

Seyed Hassan Sedighy received B.Sc., M.Sc. and PhD degrees in Electrical Engineering from Iran University of Science and Technology (IUST) in 2006 and 2008, and 2012, respectively. He is currently an associate professor with the School of advanced technologies, Iran University of Sciences and Technology. His current research interests include microstrips antenna, optical transformation, design and application of metamaterials, and RF radio links.

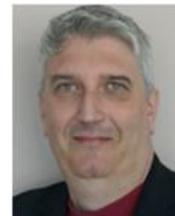

Ladislau Matekovits, (M'94–SM'11) received the degree in electronic engineering from Institutul Politehnic din Bucureşti, Bucureşti, Romania, and the Ph.D. degree (Dottorato di Ricerca) in electronic engineering from Politecnico di Torino, Torino, Italy, in 1992 and 1995, respectively. Since 1995, he has been with the Department of Electronics and Telecommunications, Politecnico di Torino, first with a post-doctoral fellowship, then as a Research Assistant. He joined the same Department as Assistant Professor in 2002 and was appointed as Senior Assistant Professor in 2005 and as Associate Professor in 2014 respectively. In February 2017 he obtained the Full Professor qualification (Italy). In late 2005, he was Visiting Scientist at the Antennas and Scattering Department, FGAN-FHR (now Fraunhofer Institute), Wachtberg, Germany. Beginning July 1, 2009, for two years he has been a Marie Curie Fellow at Macquarie University,

Sydney, NSW, Australia, where in 2013 he also held a Visiting Academic position and in 2014 has been appointed as Honorary Fellow. Since 2020 he is Honorary Professor at Polytechnic University of Timisoara, Romania and Associate of the Italian National Research Council. He has been appointed as Member of the National Council for the Attestation of University Degrees, Diplomas and Certificates (CNATDCU), Romania, for the term 2020-2024.

His main research activities concern numerical analysis of printed antennas and in particular development of new, numerically efficient full-wave techniques to analyse large arrays, and active and passive metamaterials for cloaking applications. Material parameter retrieval of these structures by inverse methods and different optimization techniques has also been considered. In the last years, bio-electromagnetic aspects have also been contemplated, as for example design of implantable antennas or development of nano-antennas for example for drug delivery applications.

He has published 375+ papers, including 90+ journal contributions, and delivered seminars on these topics all around the world: Europe, USA (AFRL/MIT-Boston), Australia, China, Russia, etc.. Prof. Matekovits has been invited to serve as Research Grant Assessor for government funding calls (Romania, Italy, Croatia and Kazakhstan) and as International Expert in PhD thesis evaluation by several Universities from Australia, India, Pakistan, Spain, etc.

Prof. Matekovits has been a recipient of various awards in international conferences, including the 1998 URSI Young Scientist Award (Thessaloniki, Greece), the Barzilai Award 1998 (young Scientist Award, granted every two years by the Italian National Electromagnetic Group), and the Best AP2000 Oral Paper on Antennas, ESA-EUREL Millennium Conference on Antennas and Propagation (Davos, Switzerland). He is recipient of the Motohisa Kanda Award 2018, for the most cited paper of the IEEE Transactions on EMC in the past five years, and more recently he has been awarded with the 2019 American Romanian Academy of Arts and Sciences (ARA) Medal of Excellence in Science and by the Ad Astra Award 2020, Senior researcher, for Excellence in Research.

He has been Assistant Chairman and Publication Chairman of the European Microwave Week 2002 (Milan, Italy), and General Chair of the 11th International Conference on Body Area Networks (BodyNets) 2016. Since 2010 he is member of the organizing committee of the International Conference on Electromagnetics in Advanced Applications (ICEAA) and he is member of the technical program committees of several conferences. He serves as Associated Editor of the IEEE ACCESS, IEEE Antennas and Wireless Propagation Letters and IET MAP and reviewer for different journals.